# EELA Operations: A standalone regional dashboard implementation

Cyril L'Orphelin, Hélène Cordier, Sylvain Reynaud
Marcos Lins, Sinikka Loikkanen, Olivier Lequeux, Pierre Veyre

*IN2P3-Computing Centre. IN2P3/CNRS*
*helene.cordier@in2p3.fr, cyril.lorphelin@in2p3.fr*

## Abstract

*Grid operators in EGEE use a dedicated dashboard as their central operational tool, stable and scalable for the last 5 years despite continuous upgrade from specifications by users, monitoring tools or data providers. In EGEE-III, regionalisation of operations led the tool developers to conceive a standalone instance of this tool. Hereby, we will present the concept and the EELA-II implementation. Indeed, the re-engineering of this tool led to an easily deployable package customized to EELA-II specific information sources such as EVENTUM, EOC DB and SAM EELA instance. This package is composed of a generic and scalable data access mechanism, Lavoisier; a widely spread php framework Symfony for configuration flexibility and a MySQL database.*

## 1. Introduction

### 1.1. General purpose of the Operations Portal

The need of a management and operations tool for EGEE and WLCG (Worldwide LCG) lead in 2004 to the creation of the EGEE Operations Portal, later referred to as "the CIC Portal"[1]. Its main focus relies on providing an entry point for all EGEE actors for their operational needs. Namely, it enables the monitoring and the daily operations of the grid resources and services through a set of synoptic views.

The variety and the distribution of tools and people involved inferred that the portal had to be an integration platform allowing strong interaction among existing tools with similar scope and filling up gaps wherever some functionalities were lacking. The biggest challenge turned out to be the size of the production infrastructure to manage and operate - over 250 sites today. At the same time comparable needs were expressed by people and bodies managing Virtual Organizations (VO), Regional Operations Centres (ROC) or Resources Centres. IN2P3 Computing Centre induced the implementation of such an integration platform, and a first prototype came out in November 2004.

### 1.2. A need for daily operations: the Operations Dashboard

The infrastructure was managed centrally from the Operations Centre at CERN at first. While this worked quite well, troubleshooting of such a large network was a hardship, and the expertise was concentrated in one place. To lessen the work load and to make sure experience with Grid operations was more evenly spread out; EGEE came up with a scheme where federations of countries shared the load. Dubbed CIC-On-Duty (COD), this new system began in October 2004 [2]. In this system, responsibility for managing the infrastructure was passing around the globe on a weekly basis.

Splitting this management responsibility reduced the workload. However, requirements on synchronization of tools and communication needs soared along with the complexity of the work. It then appeared necessary to have all the tools available through a single interface enabling a strong interactive use of these tools. The conception of this interface turned out to materialize as a communication platform between operators to become one of the main features of the Operations portal dedicated to daily operations, the COD dashboard.

This collaborative integration work turned out as a web based overview of the state of the infrastructure, used by all the teams in their daily work through their shift. This specific tool aggregates results from different monitoring tools and triggers alarms in a synoptic dashboard from which the teams have a global view of the problematic resources or grid core services. It also enables to check the administrative status versus the dynamic status of the sites. Together with web-service access to the global ticketing system, the COD dashboard brings up also to the attention of operators on duty the unattended problems at resource centres. The use of a broadcast tool for communication between all the actors in the project eases up the daily work of troubleshooting at production sites and the implementation of some consistently and regularly updated operational procedures ensured a "transversally-uniform" working manner.

Main result at the end of EGEE-II was that the combined set of operational procedures and tools and their constant evolution have been recognized key in stabilizing sites in operation. Since the early days of EGEE-III, the operational model evolved and the daily operations are now under the regions responsibility, even if still under the guidance of the operational procedures. The project still operates some supervision on the unattended problems at sites, urgent security matters or project wide operational troubles.

Consequently, the COD dashboard has evolved over the last two years into a GRIDOPS dashboard, enabling the daily operations at the federation level; the central layer indeed reduced at the minimal level possible. This reflects the evolution in the operational model of EGEE-III. The GRIDOPS dashboard is still in 2009 the tool for all operators on duty daily work and is fully implementing the EGEE Operational Procedures [3] to prepare the production infrastructure of EGEE-III to be sustainable when the operational entities will be set-up at the national level by National Grid Initiatives or NGIs at the end of EGEE-III. This additional constraint to provide an interoperable backend for distribution to existing federations, also at the national level in the next future, lead us to restructure the EGEE Operations Portal [4], to be able to provide a scalable model applicable on every level: national, federal or central.

## 2. GRIDOPS dashboard

We recently revisited the architecture of the dashboard namely for adaptability and scalability reasons and we expose below the architecture we have adopted. In section3, we will describe the EELA use-case as one of the first implementation of our recent code remanufacturing.

### 2.1. GRIDOPS dashboard Concept

We expose in Figure 1 the GRIDOPS dashboard concept, which is still an integration tool and at the same time a flexible tool deployable at the regional or at the central level. The new challenge of regionalisation induced some architecture evolution detailed in section 2.2 but the main concept created at the time of the COD dashboard remains. In fact, the aim of accessing various tools from a single entry point is a constant in our effort as the synoptic view and single operations platform proved invaluable gain for operators.

The dashboard deals with different interfaces to the source core services, as the ticketing system, the monitoring framework, or the central database containing information about sites and services as we will see in more details in section 3 dedicated to the EELA implementation.

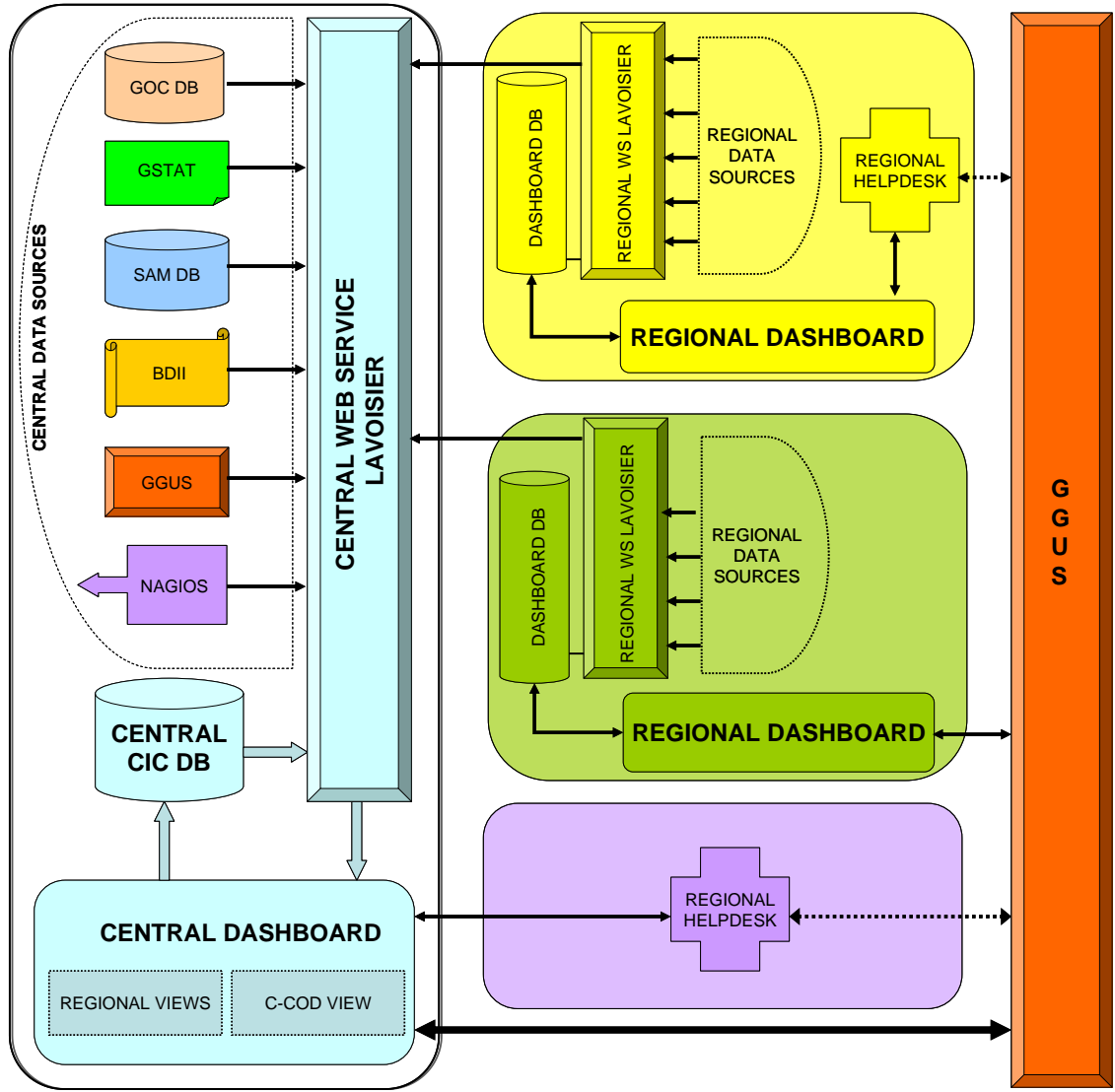

**Figure 1. Gridops dashboard concept**

Namely, the interface with EGEE central ticketing system (Global Grid User Support, GGUS [5]), and the interface with EGEE monitoring framework (Service Availability Monitor, SAM [6]), provide the operators at the regional or the central level an entry point to:
- detect problems through the gridops dashboard, browsing alarms triggered by the monitoring tool – Service Availability Monitoring SAM [6] since 2004 and NAGIOS [7] starting December 2009.
- create a ticket in EGEE ticketing system, GGUS [5], for a given problem, link this ticket to the corresponding SAM alarm, and notify responsible people using preformatted e-mails, addresses being retrieved from the Grid Operations Centre database (GOC database [8]);

- browse, modify, and escalate tickets from the GRIDOPS dashboard directly.

## 2.2. GRIDOPS dashboard architecture

The range of tools proposed in the EGEE Operations Portal is large and induces more and more complexity in related development and maintenance. To face this constant evolution, the global architecture of the portal is based on a modular structure comprising a web interface based on a PHP Framework, Symfony [9], a database and a Data Processing System named Lavoisier [10,11]. The structure has been clearly separated in different modules, enabling code factorization and reusability as well as an ease in portal deployment. We will focus in this section on the Web Portal design and the Lavoisier module in section 2.3.

The Web Portal component is mainly re-written within a PHP Framework; it represents the user interface. It is based on a View-Controller pattern and all the user requests are filtered out and checked by the main controller. Whenever authorisation is needed, authentication is done using X509 certificate. Accepted requests are then forwarded to the sub-controller in charge of loading the requested page. In addition, the main controller uses abstraction layers to transparently connect to third-party applications like databases or web services. Connections are shared by all the sub-controllers thanks to inheritance mechanisms. This approach improves organisation of source code, leading to a high rate of reusability. Moreover, all the operating system dependent implementations have been removed in order to ease configuration and deployment. It means also that the dashboard will be provided with a data schema and that with the help of the Symfony Framework you can generate on the fly the Database corresponding to the schema. This option is working with MySQL, Oracle, PostgressSQL and SQLLite. The Symfony Framework is also giving additional features like a security layer assuming XSS and CSRF Protection, a set of different work environments different: test, prod and dev, and the use of plug-ins developped by the Symfony community.

As a conclusion, this framework will ease the configurability, maintenance end extension of the functions of the dashboard functionality, one of the main features of the operations portal. The architecture has been designed to be efficient in the integration of multiple data sources and the modularity and the reusability of code in Symfony ease this integration. But the key component is the Data Source Composition Service Lavoisier.

## 2.3. General Workflow - Lavoisier

We are highly dependent from the evolution of other tools: GOC DB, GGUS, SAM, Nagios (see figure 1) and our technical solution is organized around a web service implementation, to make the integration of these changes transparent. Consequently, the main idea is to have the integration of each resource via Lavoisier.

"Lavoisier" [10, 11] has been developed in order to reduce the complexity induced by the various technologies, protocols and data formats used by its data sources. It is an extensible service for providing a unified view of data collected from multiple heterogeneous data sources. It enables to easily and efficiently execute cross data sources queries, independently of used technologies. Data views are represented as XML documents [12] and the query language is XSL [13].

The design of Lavoisier enables a clear separation in three roles: the *client* role, the *service administrator* role and the *adapter developer* role. The main *client* is the CIC portal; it can retrieve the content of a data view in XML or JSON format through SOAP or REST requests. It can also submit XSL style sheets that will be processed by Lavoisier on the managed data views, and receive the result of this processing.

The *service administrator* is responsible for configuring each data view. He must configure the adapter, which will generate the XML data view from the legacy data source. He may also configure a data cache management policy, in order to optimize Lavoisier according to the characteristics and the usage profile of both the data source (e.g. amount of data transferred to build the view, update frequency, latency) and the generated data view (e.g. amount of data, time to live of the content, tolerable latency). Data cache management policy configuration includes:

- the cache type (in-memory DOM tree, on-disk XML file/files tree or no cache),
- view dependencies,
- a set of rules for triggering cache updates, depending on time-based events, notification events, data view read or write access, cache expiration, update of a cache dependency, etc.
  a set of fallback rules for ignoring, raising errors or retrying cache updates in case of failure, depending on the type of the exception thrown,
- the cache time-to-live, to prevent from providing outdated information in case of successive cache update failures,
- a cache update timeout,
  the synchronization of the exposition of the new cache content for inter-dependant data views, in order to ensure data view consistency,
- the validation mode for generated views (conform to XML Schema [14], well-formed XML or no validation).

The configuration is re-loadable on the fly. Then only the reconfigured data views and their dependencies are suspended during the configuration reload. This enables the service administrator to add, remove or reconfigure data views with a minimal service interruption. Moreover, in most cases, configuration changes will have no impact on the code of the data consuming applications and of the adapters.

The *adapter developer* adds to Lavoisier the support for new data sources technologies by implementing a set of required and optional interfaces. Some reusable adapters are provided to access data using various technologies, such as RDBMS, LDAP, Web Services, JMS, XML command line output stream, local and remote flat, XML or HTML files, etc. Other reusable adapters take an existing data view and transform it to another data view, using technologies such as XSL, XQuery, SAX-based XML filtering. Introspection adapters expose data about data views configuration and current cache state.

Figure 2 illustrates a very simple example of Lavoisier configuration with four data views. Data views obtained from flat file and RDBMS data sources are cached respectively in memory and on disk, while the data view obtained from the Web Service data source is regenerated each time it is accessed. The fourth data view is generated from the RDBMS data view, and refreshing of its in-memory cache is triggered when the cache of the RDBMS data view is refreshed. These two data views can be configured to exposed their new cache simultaneously if consistency is required.

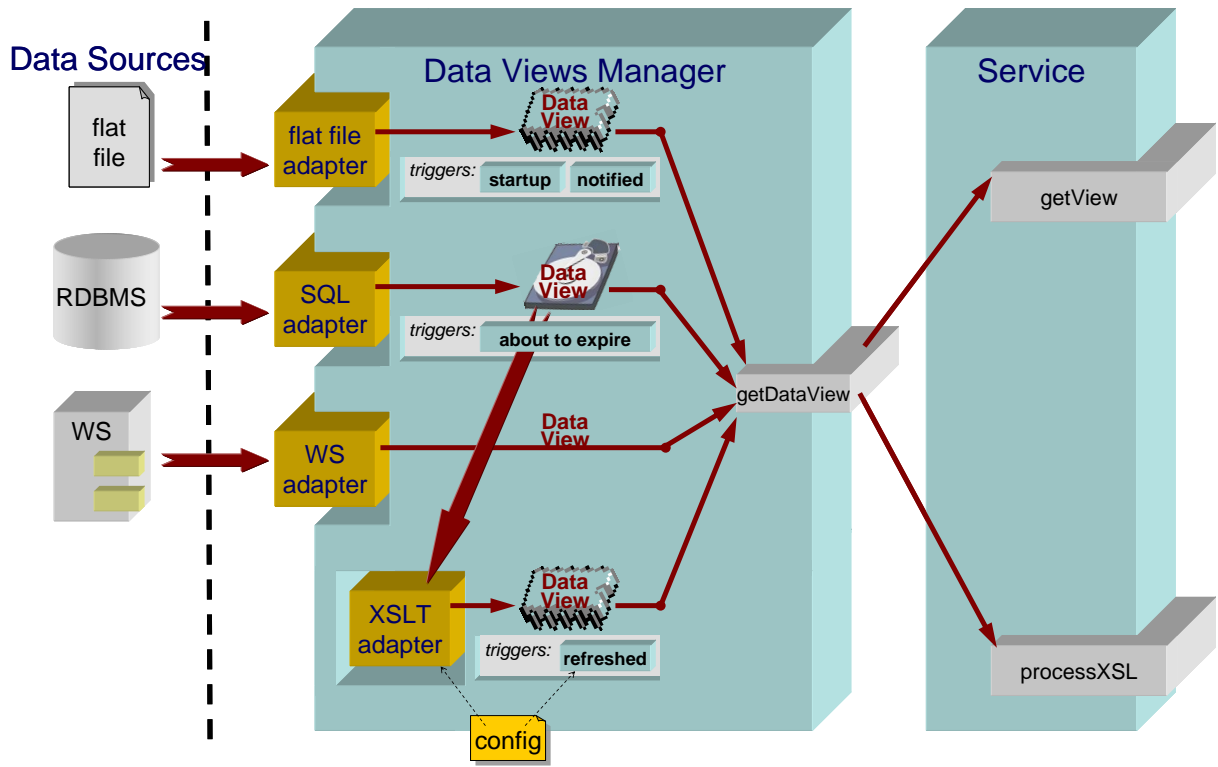

**Figure 2. Data Source Composition Service Lavoisier Configuration**

Lavoisier has proven effective in increasing maintainability of the CIC portal, by making its code independent from the technologies used by the data sources and from the data cache management policy. Its design and interfaces made easier writing reusable code, and good performances are easily obtained by tuning the cache mechanisms in an absolutely transparent way for the portal code. Indeed, the different components work in a standardized way through the output of the Lavoisier Web Service. The translation of resource information into this standardized output is provided by different plug-ins.

Recently, a huge joint effort with EELA-II has been recently put into the configuration of Lavoisier, the structure of its caches and the rules of refreshing to have efficient, scalable and reliable data handling. Indeed, all information has been structured around the base component of the Operational Model: the site. We retrieve the global information about primary sources like GGUS, GOC DB, SAM and we organize it by sites. The main idea is to construct a summary of the different available information for a site: firstly, this organization permits to continue to work with the caches, even if a primary source is unavailable; then you access only information you need on the web page. Your information is structured around a synoptic view of the site and you don't access hundreds of times the primary sources but a subset of them with the site view. Finally, we refresh the data sources only when we need it and only when an action has been triggered. Last but not least, it is very easy to add a new data source in this model. In the configuration file of Lavoisier you add the access to the primary source and also the split of this information per site. This information is then readily available in the synoptic view of the site.

## 3. EELA specific implementation

Given our recent restructuration of the GRIDOPS dashboard concept according to figure 1 and to the architecture enhancements, initiated as a joint collaboration between EELA-II and CC-IN2P3 involving Lavoisier component, we are about to deliver in a matter of weeks a dashboard for operations dedicated to EELA operational needs. As we have exposed the Symfony framework as well as the role and principles of Lavoisier, we will expose below the package configuration and then the data sources relevant to EELA customization.

### 3.1. Installation Requirements

EELA-II will be able to install a GRIDOPS dashboard customized to its needs with a minimum set of requirements. Indeed the EELA-GRIDOPS server will have to run PHP with a version over 5.2.4 and some modules enabled suck as SOAP and OCI, together with Java with a JDK version over 5.0 and a database of its choice MySQL, SQLLite, Oracle or PostgresSQL. The full package delivered will comprise the PHP code, the Database schema and the Lavoisier module, together with the relevant configuration files to tune the GRIDOPS behavior to EELA operational needs.

### 3.2. Configuration of information sources

The interactions of the GRIDOPS dashboard with the EELA specific information sources when possible for each source type: monitoring source (SAM [15]), helpdesk (Eventum [16]) and static repositories (EOC [17]) are summarized in Figure 3.

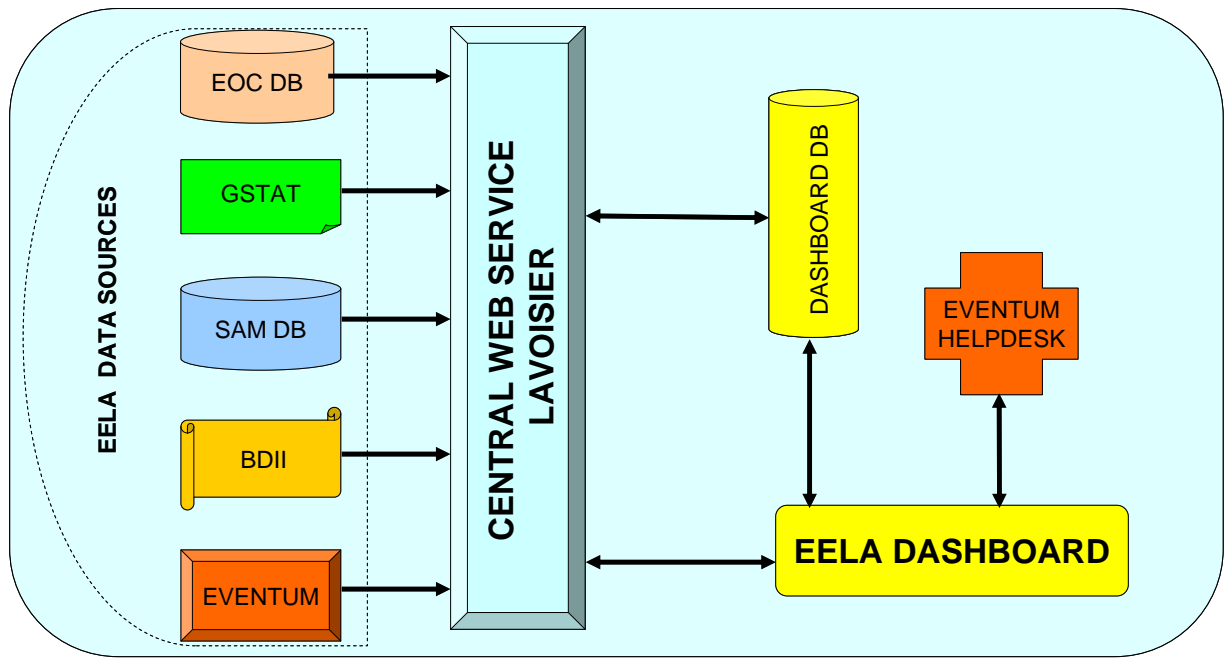

**Figure 3. The EELA specific information sources**

#### 3.2.1. Monitoring

SAM, is a framework providing sensors, metrics and alarms for services in EGEE-WLCG infrastructure developed at CERN [6]. EELA has a dedicated SAM instance in Brazil [15] so the workflow is the same: Sensor tests are regularly submitted and each failure triggers the creation of an alarm entry in a database, giving details about sensor, test, node, date of failure and so on. These alarms are used by operators as starting point to detect and report problems. Consequently, the list of new alarms will appear on the GRIDOPS dashboard, and an interface is needed to link each alarm to the test information on one side, and to the ticket creation process on the other side. The workflow that deals with the SAM alarm treatment in exposed in figure 4.

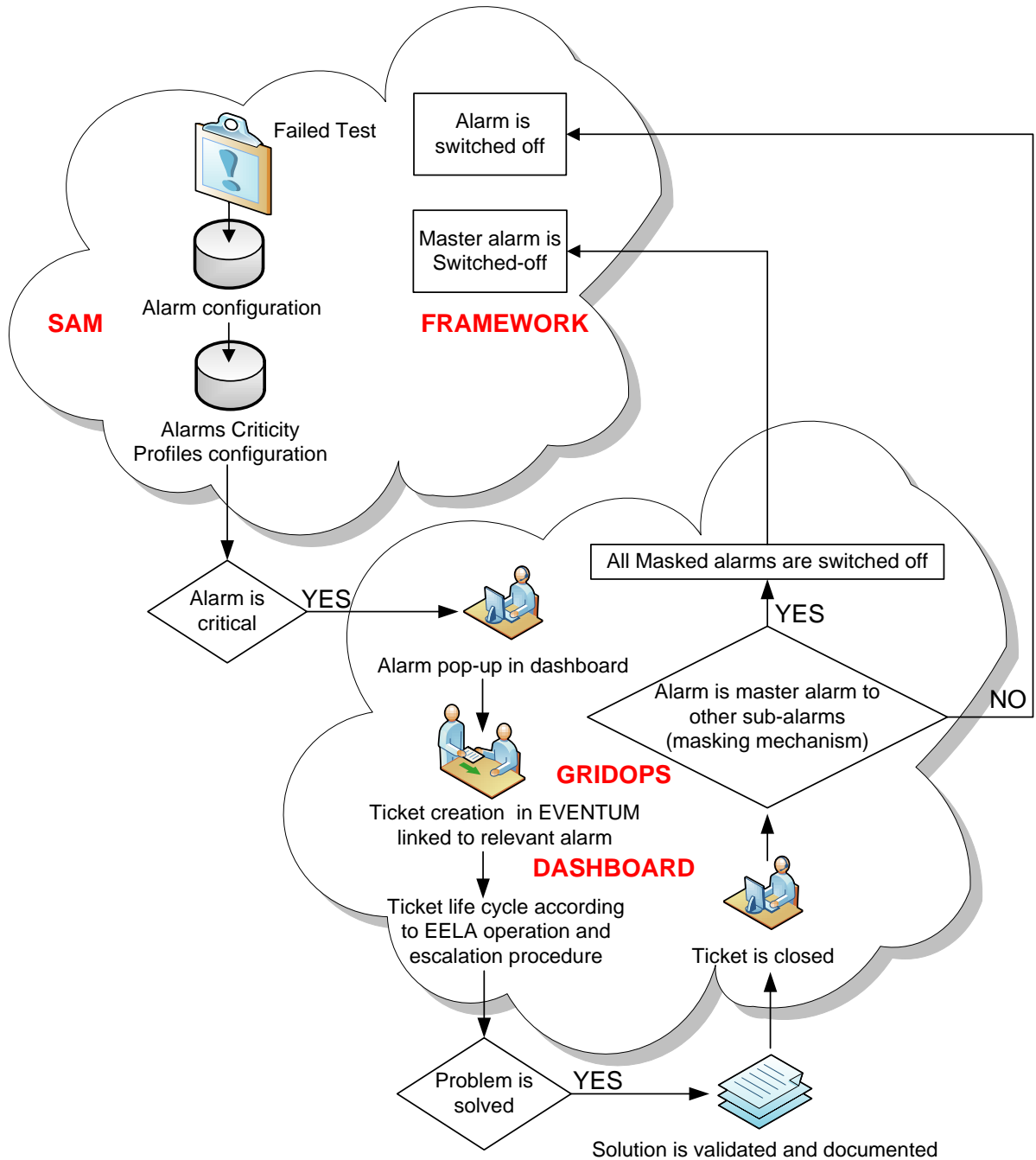

**Figure 4. SAM alarm workflow**

Indeed, when created in SAM DB, a new alarm is shown on the GRIDOPS dashboard along with all the useful information allowing operators to provide diagnosis and to act accordingly: either reporting and tracking the problem, or setting off the alarm if the failure appears to be temporary. Alarms handling includes the notion of alarm status: alarms can be masked one by another, or designed as "assigned" if they are already taken into account. Consequently, the GRIDOPS dashboard is the tool to detect problems, to browse alarms triggered by SAM, to create a ticket in relevant ticketing system, to link this ticket to the corresponding SAM alarm, and to notify responsible people using preformatted e-mails, addresses being retrieved from static repository used by EELA and finally to browse, modify, escalate or close tickets according to EELA procedures.

### 3.2.2. Helpdesk

GGUS is the official User Support system within EGEE and WLCG and its core mechanism consists of a ticketing system based on Remedy [18]. As mentioned above, the EGEE Operations Portal is used for daily operations support. Grid Operators have to monitor grid resources using available monitoring tools and track problems using GGUS. To provide a single entry point for operators, the front-end and back-end are decoupled: with the core mechanism hosted by GGUS, and the user interface hosted on the CIC portal.

The back-end consists of a dedicated database in the GGUS Remedy system called "CIC_Helpdesk": the central Helpdesk could not directly be used because of specific format of operations tickets (special fields for implied site and for impacted node, and different escalation steps).

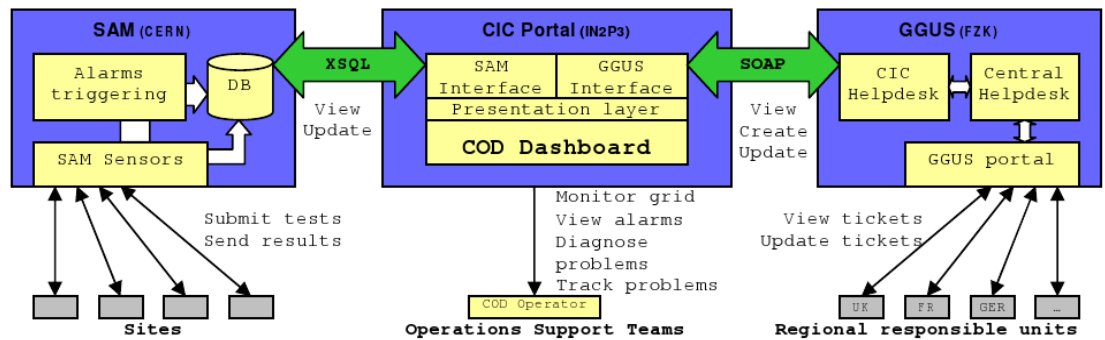

**Figure 5. SAM and GGUS Interfaces to the COD and GRIDOPS Dashboard**

When assigned to a given Responsible Unit, the CIC_Helpdesk ticket is duplicated in the central Helpdesk where it is treated as all other tickets. Any change on one of the two tickets triggers the same modification on the other to ensure synchronization (see Figure 5).

The interface appears in the COD dashboard: tickets are listed, created, modified and escalated from web pages, where this information is coupled with monitoring results.

This interface, mainly written in PHP, communicates with the back-end via SOAP web services which allow performing all operations on tickets as illustrated on Figure 5.

The mechanism is the same for the new GRIDOPS dashboard: direct actions such as ticket creation or update are performed using SOAP messages encapsulated with the SOAP module of PHP5, while tickets list is built using Lavoisier. The interface with the GRIDOPS dashboard communicates with SAM back-end using a XSQL-based service and direct actions on alarms are performed directly querying the service. Lists of alarms are built using Lavoisier. Finally, a data view is built grouping information from GGUS and from SAM.

The global workflow will be the same with Eventum but the connections will be different. Firstly, we will connect directly to a Mysql Database and we will create ticket by direct insertions in the Eventum DB. Some questions still remain though on how to integrate specific operational steps and how we can translate it into the GRIDOPS dashboard.

### 3.2.3. Static repositories

Our solution has been designed to work with the GOCDB Programmatic Interface. The GOC DB is providing methods to access to the data related to site in a standard way base on XML Files. A solution based on the integration of EOC DB with the GOC PI is under study between EOC DB and GOC DB representatives. In this case the integration of EELA sites will be immediate. Otherwise we will retrieve information in Lavoisier and organize it to have the same structure of files we use for information coming from GOC DB. The integration on the web part will be the same.

## 4. Conclusions

EGEE COD dashboard has evolved into GRIDOPS dashboard coping with the new EGEE - III operations model. From December 2009 on, federations will be able to operate a centralized "view" or a standalone instance of the dashboard in line with the project procedures operating the global production infrastructure. Indeed, a central dashboard will supervise without disruption for the EGEE-III project any of the two types of dashboard that the federations will adopt for their daily operations.

We showed that the current EGEE Operations Model induced a drastic transformation in the dashboard backend structure and configuration files to cope with the underlying challenges concerning flexibility, scalability and reliability. Consequently, with this code restructuration, we can improve easily our tool with external partners in order to meet requirements from any future operational entities like grid projects, federations or nations.

Future work comprises similar restructuration of the other features of the EGEE Operations Portal. Our goal is to provide an easy access to administration interfaces for sites or VO managers, a standard access to information and standard formats, together with a scalable model for each level: national, federal or central. Moreover, for user ergonomy and backend simplicity we are working on common development plans with GOCDB [19].

As a conclusion, the GRIDOPS dashboard configurable to EELA purposes is the first use-case of early adoption of this highly flexible integration platform outside EGEE. It followed in the past the evolution of the operational model of EGEE, but we are capable now to provide a generic tool able to follow easily all kind of procedure evolution without much effort. Indeed, enforcing EELA operational model in the coming weeks will be a very sound proof of concept.

## 5. References

[1] (2004) CIC Operations portal [online]: *http://cic.gridops.org*

[2] H. Cordier, G. Mathieu, F. Schaer, J. Novak, P. Nyczyk, M. Schulz, M.H. Tsai*, Grid Operations: the evolution of operational model over the first year,* Computing in High Energy and Nuclear Physics, India, 2006.

[3] EGEE Operational procedures https://edms.cern.ch/document/840932

[4] CIC portal: a Collaborative and scalable integration platform for high availability grid operations: Aidel, O., Cavalli, A., Cordier, H., L'Orphelin, C., Mathieu, G., Pagano, A., Reynaud, S., Grid Computing, 2007 8th IEEE/ACM Int. Conf; USA. 2007 Page(s):121 - 128

[5] GGUS Web Portal [on line] http://www.ggus.org

[6] SAM wiki http://goc.grid.sinica.edu.tw/gocwiki/Service_Availability_Monitoring_Environment

[7] NAGIOS [on line] http://www.nagios.org/docs/

[8] GOC portal [online]: http://goc.grid-support.ac.uk

[9] Symfony http://www.symfony-project.org/doc/1_2/

[10] S. Reynaud, G. Mathieu, P. Girard, F. Hernandez and O. Aidel, *Lavoisier: A Data Aggregation And Unification Service*, Computing in High Energy and Nuclear Physics, India, 2006.

[11]   Lavoisier documentation [online]. Available: http://grid.in2p3.fr/lavoisier

[12]   (1996-2003) Extensible Markup Language (XML) [online] http://www.w3.org/XML

[13]   (1999) XSL transformations [online] : http://www.w3.org/TR /xslt

[14]   http://www.w3.org/XML/Schema

[15]   EELA SAM instance http://sam.eela.ufrj.br/

[16]   EVENTUM http://eventum.eu-eela.eu

[17]   EOC http://eoc.eu-eela.eu/doku.php?id=quick_guide

[18]   (2005-2007) BMC Software - Remedy system [on line]: http://www.bmc.com/remedy/

[19]   Poster Session: EGEE Operations Portal: From an Integration Platform into a Generic Framework in EGI context, EGEE'09, Spain, 2009.